\documentclass[%
 aip,
 jmp,%
 amsmath,amssymb,
 reprint,%
]{revtex4-1}

\usepackage{subfigure}% Include figure files
\usepackage{graphicx}% Include figure files
\graphicspath{{./}{./images/}{./images_all/}{./images_all/single_magnet_computing/}}
\DeclareGraphicsExtensions{.eps}
\usepackage{dcolumn}% Align table columns on decimal point
\usepackage{bm}% bold math
\begin{document}

\preprint{AIP/123-QED}

\title{Ultra-low-energy non-volatile straintronic computing using single multiferroic composites}% Force line breaks with \\

\author{Kuntal Roy}
\email{royk@vcu.edu.}
\noaffiliation
\affiliation{School of Electrical and Computer Engineering, Virginia Commonwealth University, Richmond, VA 23284, USA}
\thanks{Current affiliation: School of Electrical and Computer Engineering, Purdue University, West Lafayette, IN 47907, USA.}

%\date{\today}% It is always \today, today,
             %  but any date may be explicitly specified

\begin{abstract}
The primary impediment to continued downscaling of traditional charge-based electronic devices in accordance with Moore's law is the excessive energy dissipation that takes place in the device during switching of bits. One very promising solution is to utilize multiferroic heterostructures, comprised of a single-domain magnetostrictive nanomagnet strain-coupled to a piezoelectric layer, in which the magnetization can be switched between its two stable states while dissipating minuscule amount of energy. However, no efficient and viable means of computing is proposed so far. Here we show that such single multiferroic composites can act as universal logic gates for computing purposes, which we demonstrate by solving the stochastic Landau-Lifshitz-Gilbert (LLG) equation of magnetization dynamics in the presence of room-temperature thermal fluctuations. The proposed concept can overwhelmingly simplify the design of large-scale circuits and portend a highly dense yet an ultra-low-energy computing paradigm for our future information processing systems.
\end{abstract}

\maketitle

Utilizing electron's spin rather than its charge as state variable has been widely studied in the field of so-called spintronics,~\cite{RefWorks:89} particularly in the context of nanomagnets,~\cite{RefWorks:132,RefWorks:328} since it can potentially lead to ultra-low-energy computing. Recently, it has been shown that the magnetization of a 2-phase multiferroic composites,~\cite{RefWorks:164} comprised of a single-domain magnetostrictive nanomagnet strain-coupled to a piezoelectric layer, can be switched between its two stable states with a tiny voltage of few tens of millivolts at room-temperature.~\cite{roy11,roy11_6,roy13_spin} Such electric-field induced magnetization switching mechanism dissipates a miniscule amount of energy of only $\sim$1 attojoule with sub-nanosecond switching delay at room-temperature~\cite{roy11_6} and thus it can potentially extend the lifeline of conventional electronics.~\cite{moore65,RefWorks:553,RefWorks:211} Experimental efforts to demonstrate the operation of such straintronic devices are emerging too.~\cite{RefWorks:551,RefWorks:559,RefWorks:611,RefWorks:609}

Here, we propose a viable and an efficient way of devising logic elements exploiting such devices for general-purpose computing. With experimentally feasible parameters, we theoretically demonstrate that such \emph{single} multiferroic elements can act as universal logic gates. Traditionally, implementing logic gates according to magnetic quantum cellular automata (MQCA) architecture~\cite{RefWorks:134,RefWorks:154,roy13_spin,fasha11} takes multiple elements to implement the same logic functionality and thus incurs more complexity, switching delay, energy dissipation, and area on a chip. Therefore, the proposed concept has profound promise in simplifying the design of large scale circuits and consequently improving the performance metrics drastically.

\begin{figure}[b]
\centering
\includegraphics[width=80mm]{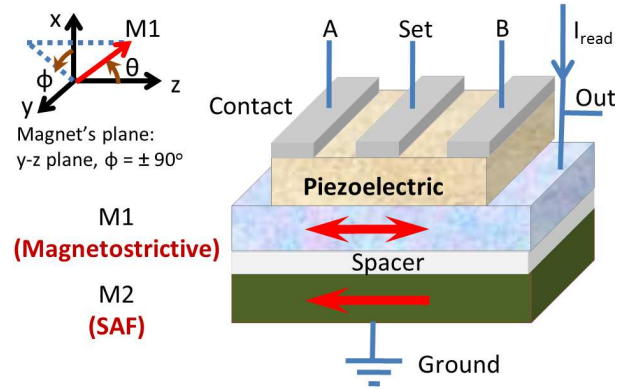}
\caption{\label{fig:computing_single_magnetostrictive_diagram} {Schematics of the proposed single-element straintronic universal logic gates.} 
By applying voltages at the terminals $A$ and $B$, the magnetization of the magnetostrictive nanomagnet (layer M1) can be switched. The \emph{spacer} layer is a thin layer ($\sim$1 nanometer) made of materials like Magnesium Oxide (MgO) for tunneling magnetoresistance (TMR) measurement in magnetic tunnel junction (MTJ) structures (see Ref.~\onlinecite{RefWorks:33}). The M2 layer is a synthetic antiferromagnetic (SAF) structure (see Ref.~\onlinecite{RefWorks:300}) and is permanently magnetized along one of two orientations along its easy axis, say the -$z$-axis. The output of the gate is extracted from the magnetoresistance (MR) measurement of the MTJ structure (layers M1 and M2 separated by the spacer) by passing a current $I_{read}$. The $Set$ terminal is required to set the magnetization direction of the M1 layer opposite to that of the M2 layer.} 
\end{figure}

\begin{figure*}[t]
\centering
\includegraphics[width=180mm]{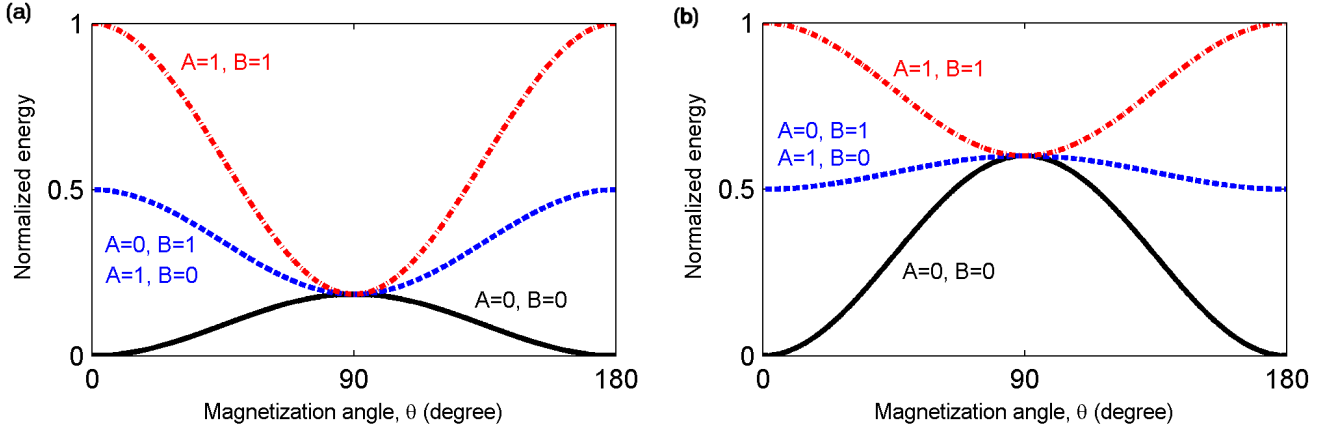}
\caption{\label{fig:potential_profile} {Potential landscapes of the magnetostrictive nanomagnet (M1 layer) with different input vector combinations as a function of the angle $\theta$ subtended by its magnetization vector with the +$z$-axis for NOR and NAND logic operations, respectively.} The potential profiles are shown when the magnetization vector lies on the magnet's plane, i.e., $y$-$z$ plane. When no voltage is applied to either of the inputs $A$ and $B$, the potential landscape is the shape-anisotropic energy barrier of the nanomagnet. (a) Potential landscapes corresponding to the NOR gate. When voltage is applied to either one of the inputs or both, the generated stress-anisotropy inverts the potential landscape and switching takes place. (b) Potential landscapes corresponding to the NAND gate. In this case, only when voltages are applied to both the inputs, the generated stress-anisotropy can invert the potential landscape and switching can take place.}
\end{figure*}

Although a number of proposals have been reported on spintronics for the purposes of computing,~\cite{RefWorks:443,RefWorks:123,RefWorks:290,RefWorks:289,RefWorks:180} the present proposal with single multiferroic composite structures is \emph{unique} in the sense that it \emph{simultaneously} satisfies the following important attributes of \emph{general-purpose} computing: ultra-low-energy dissipation, fast (sub-nanosecond) switching, room-temperature operation, and highly dense logical functionality per unit area. Using single multiferroic elements as universal logic gates while simultaneously being highly energy-efficient would facilitate to cram more functionality on a chip and hence it has immense potential to be an important contributor to Beyond Moore's law technologies.~\cite{moore65,RefWorks:211}

The basic structure of the proposed universal logic gates using single multiferroic elements is shown in Fig.~\ref{fig:computing_single_magnetostrictive_diagram}. Application of voltages at the input terminals $A$ and $B$ generates strain in the piezoelectric layer (two inputs generate \emph{twice} as much strain compared to when voltage is applied to only one input) and the strain is transferred elastically to the magnetostrictive nanomagnet (M1 layer).~\cite{roy11} This generates a stress-anisotropy that can overcome the shape-anisotropy of the nanomagnet M1 to switch its magnetization (LOGIC operation, to be described later).~\cite{roy11,RefWorks:158} The magnetization direction of the M1 layer can be switched opposite to that of the M2 layer (with fixed magnetization direction) by application of a voltage at the $Set$ terminal on the piezoelectric layer. This is termed as SET operation, which is required to perform before a LOGIC write operation, however, once the bit is written, the logic output can be read (READ operation) as many times as required before any further write operation. The output of the gate (the $Out$ terminal) is extracted from the read-line measuring magnetoresistance (MR),~\cite{RefWorks:33,RefWorks:300} of the structure i.e., if the relative orientation of the magnetizations in the layers M1 and M2 is parallel, MR is \emph{low} and the output is \emph{logic 0}, while for the anti-parallel case, MR is \emph{high} and the output is \emph{logic 1}. Since the output from a gate is connected to the inputs of the gates on the next stage and the inputs are on a \emph{thick} piezoelectric layer (so it does not load the output much), the \emph{universal} logic gates can be concatenated to achieve any Boolean logic function.

Figures~\ref{fig:potential_profile}a and~\ref{fig:potential_profile}b show the potential profiles of the magnetostrictive nanomagnet (layer M1 in the Fig.~\ref{fig:computing_single_magnetostrictive_diagram}) for different input vector combinations corresponding to NOR and NAND logic operations, respectively. The plots depict that the potential landscapes of the magnetostrictive nanomagnets can be inverted with application of voltages at the terminals $A$ and $B$ generating stress-anisotropy~\cite{RefWorks:158,roy11} in the nanomagnet, so that the minimum energy position changes from nanomagnet's easy-axis ($\theta=0^\circ$ or $180^\circ$) to its hard-axis ($\theta=90^\circ$). The output \emph{logic 0} actually corresponds to some finite voltage due to small but non-zero resistance of the MTJ, however, such small voltages are not enough to invert the barrier and enable switching.

\begin{figure}
\centering
\includegraphics[width=80mm]{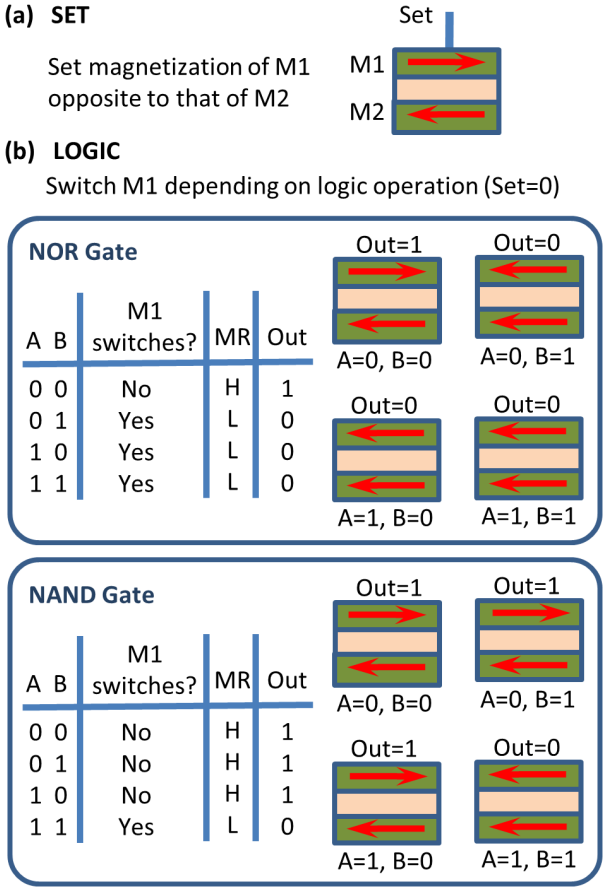}
\caption{\label{fig:computing_single_magnetostrictive_gates} {Operations of straintronic universal logic gates.} Magnetoresistance (MR) is \emph{low} (L) [\emph{high} (H)] depending on the parallel [anti-parallel] orientation of the magnetizations in the layers M1 and M2. (a) The SET operation switches the magnetization of the M1 layer to the opposite to that of the M2 layer. This operation should precede the LOGIC write operation in the M1 layer. (b) LOGIC operations of the universal logic gates NOR and NAND after the SET operation.}
\end{figure}

The potential profiles are shown when magnetization lies on its plane, however, consideration of magnetization's dynamics in full three-dimensional space is required for a complete 180$^\circ$ switching of magnetization even in the presence of room-temperature thermal fluctuations.~\cite{roy11_5} Computing methodologies utilizing such 180$^\circ$ switching mechanism between the two stable states of a shape-anisotropic magnetostrictive nanomagnet have not been proposed so far. Note that the potential energies of the corresponding nanomagnets for the NOR and NAND gates are drawn in their respective normalized scales. The shape-anisotropic energy barriers ($A$=0, $B$=0 cases) for both the NOR and NAND gates are of same magnitude since it is a design criterion that determines the thermal stability or the error-probability due to spontaneous switching of magnetization. The nanomagnets designed for the NOR and NAND gates need to be of same thickness should both types of the gates are required on a chip simultaneously.~\cite{zhou07} Both of these universal logic gates can operate using nanomagnets with the same material and voltage level provided they are designed with different lateral dimensions (to be described later). The principles of operation of these two gates are shown in the Fig.~\ref{fig:computing_single_magnetostrictive_gates}. Basically, during LOGIC operation, depending on the stress level and the type of gate, the potential barrier of the nanomagnet M1 gets inverted and the magnetization switches, which makes the magnetoresistance \emph{low} and thus it performs the respective logic operation for the gates.

The design of the nanomagnets for devising NOR and NAND gates are different due to their respective logic operations. This can be understood from the modifications of the potential energy barriers required for the gates as depicted in the Fig.~\ref{fig:potential_profile}. The critical stress needed to overcome the unperturbed shape-anisotropic potential barrier ($A$=0, $B$=0 case) is lower for the NOR gate than that of the NAND gate. The stress-anisotropy in a magnetostrictive nanomagnet is proportional to the product of stress and volume of the nanomagnet;~\cite{RefWorks:158,roy11} hence the volume of the nanomagnet for devising the NOR gate is required to be higher than the one used for the NAND gate. Now, the shape-anisotropy energy barrier height is proportional to a nanomagnet's volume and the degree of aspect ratio of the nanomagnet's elliptical cross-section for a given thickness.~\cite{RefWorks:158,roy11} Since the nanomagnet for devising the NAND gate is of lower volume, the aspect ratio of its elliptical cross-section needs to be higher. Thus, for a given thickness, the lateral dimensions of the nanomagnet for devising the NAND gate is smaller than the one for the case of the NOR gate.~\cite{RefWorks:402} 

\begin{figure*}
\centering
\includegraphics[width=180mm]{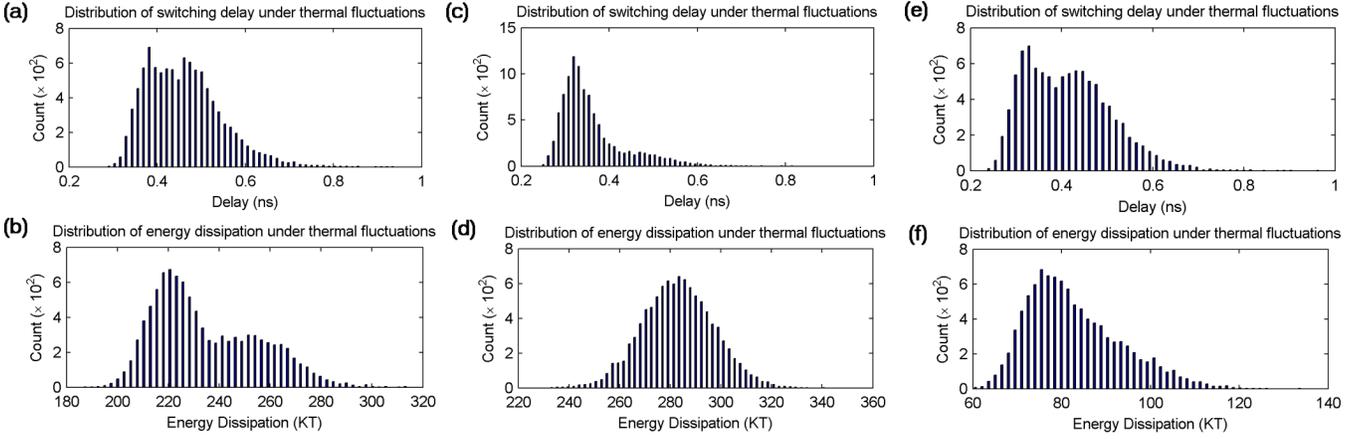}
\caption{\label{fig:results_delay_energy} Delay and energy distributions for switching of magnetization for the NOR and NAND gates at room temperature (300 K). 
(a) Distribution of the switching delay (with mean 0.46 ns and standard deviation 85 ps) and (b) Distribution of energy dissipation (with mean 234.8 $kT$ and standard deviation 20.5 $kT$) when only one input is ON (15 MPa stress) for the NOR gate. 
(c) Distribution of the switching delay (with mean 0.36 ns and standard deviation 72 ps) and (d) Distribution of energy dissipation (with mean 283.2 $kT$ and standard deviation 14.1 $kT$)when both the inputs are ON (30 MPa stress) for the NOR gate.
(e) Distribution of the switching delay (with mean 0.41 ns and standard deviation 93 ps) and (f) Distribution of energy dissipation (with mean 82.8 $kT$ and standard deviation 10.5 $kT$) when both the inputs are ON (30 MPa stress) for the NAND gate.}
\end{figure*}

We have solved stochastic Landau-Lifshitz-Gilbert (LLG) equation~\cite{RefWorks:162,RefWorks:161,RefWorks:186} to design the universal logic gates NOR and NAND in the presence of room-temperature thermal fluctuations~\cite{roy11_6}. The material parameters that characterize the magnetostrictive layer made of polycrystalline Terfenol-D (TbDyFe) are as follows -- Young's modulus (Y): 80 GPa, Magnetostrictive coefficient ($(3/2)\lambda_s$): +90$\times$10$^{-5}$, saturation magnetization ($M_s$): 8$\times$10$^5$ A/m, and Gilbert's damping parameter ($\alpha$): 0.1 (Ref.~\onlinecite{RefWorks:179,RefWorks:176,RefWorks:178,roy11_6}). We model the nanomagnets as elliptical cylinders and the dimensions of the nanomagnets for designing the NOR and NAND gates are chosen as $117\,nm \times 102\,nm \times 6\,nm$ and $70\,nm \times 52\,nm \times 6\,nm$, respectively. These dimensions ensure that the nanomagnets have a \emph{single} ferromagnetic domain.~\cite{RefWorks:133,RefWorks:402} Alongwith the material parameters, the dimensions ensure that the in-plane static shape-anisotropy energy barrier height is $\sim$60 $kT$ at room-temperature for both the gates. For the piezoelectric layer, we use lead-zirconate-titanate (PZT). We will assume that the maximum strain that can be generated in the PZT layer is 500 ppm,~\cite{RefWorks:170} which would require a voltage of 66.7 mV because $d_{31}$=1.8$\times$10$^{-10}$ m/V for PZT~\cite{roy11_6} and the PZT layer is assumed to be 24 nm thick.~\cite{roy11_6} The corresponding stress is the product of the generated strain ($500\times10^{-6}$) and the Young's modulus of the  magnetostrictive layer. So the maximum stress that can be generated on the Terfenol-D layer is 40 MPa. 

Switching delay and total energy dissipation are calculated following the prescription in Refs.~\onlinecite{roy11_2,roy11_6}. We determine the initial distributions of polar angle $\theta$ and azimuthal angle $\phi$ at room-temperature~\cite{supplx_add_sim} and we perform a moderately large number (10,000) of simulations for each value of stress and gate type to generate the simulation results in this Letter. The ramp duration of stress is assumed to be 60 ps.~\cite{roy11_6} We assume that each of the inputs $A$ and $B$ to the logic gate can generate 15 MPa stress (corresponding to voltage 25 mv) on the magnetostrictive nanomagnet so that when both the inputs are turned ON, the total stress on the nanomagnet would be 30 MPa.

The read current $I_{read}$ needs to be calculated in a way such that a gate upon concatenation to a subsequent stage can apply the same voltage of 25 mv at one input. We can determine and calibrate the MTJ resistance for both anti-parallel ($R_{AP}$) and parallel ($R_P$) condition. Then we need to apply the following equation: $25\;mv\, = I_{read} \times R_{AP}$. With $R_{AP}=25\,M\Omega$, the read current $I_{read}=1\,nA$. Assuming $R_{AP} = 10\, R_P$,~\cite{RefWorks:581} we will have only 5 mv of voltages applied when both the inputs are \emph{logic 0}; this corresponds to 3 MPa of stress on the magnetostrictive nanomagnet, which is not sufficient to switch its magnetization. Note that the application of a voltage at the $Set$ terminal to switch the magnetization of M1 layer to the opposite to that of the M2 layer would require the knowledgebase of the magnetoresistance of the MTJ. If the relative orientation is anti-parallel, no voltage is applied; otherwise, a voltage is applied to switch the magnetization of the M1 layer to make the relative orientation anti-parallel.

Figures~\ref{fig:results_delay_energy}a and~\ref{fig:results_delay_energy}b show the distributions of switching delay and energy dissipation, respectively for the NOR logic gate upon application of 15 MPa stress (only one input is ON), while Figs.~\ref{fig:results_delay_energy}c and~\ref{fig:results_delay_energy}d show the same for 30 MPa stress (when both inputs are ON). The mean energy dissipation to perform switching in a NOR LOGIC operation is less than 1.25 attoJoule (aJ) at sub-nanosecond switching delay. Figures~\ref{fig:results_delay_energy}c and~\ref{fig:results_delay_energy}d show the distributions of switching delay and energy dissipation, respectively for the NAND logic gate when both the inputs are ON (30 MPa stress). Note that when only one input is ON, it is not sufficient for switching to take place. For switching in a NAND LOGIC operation, the mean energy dissipation is around 0.35 aJ at room-temperature and the switching takes place in sub-nanosecond time-frame too. The overall mean energy dissipation can be much less if we consider switching for different input vector combinations, e.g., for NAND LOGIC operation, since switching of magnetization takes place only for one input combination ($A$=1, $B$=1), considering equal probability of different input vectors, the mean energy dissipation can be as low as 0.1 aJ at room-temperature. The SET operation also incurs similar amount of energy dissipation as of LOGIC operation when magnetization direction is required to switch. Note that with a pipeline of subsequent SET and LOGIC operations, the \emph{effective} switching period is not affected. Such ultra-low-energy magnetic logic systems can be powered by energy harvesting assemblies~\cite{roy11,roundyf, anton, RefWorks:556} that can harvest energy from the environment without the need of an external battery. 

The SET operation precedes the LOGIC operation since after LOGIC operation the anti-parallel orientation between the magnetic layers may become parallel. However, it should be noted particularly for NAND LOGIC operation, the magnetic orientation may become parallel only if both the inputs are \emph{logic 1}, while for the other three input combinations, the magnetic orientation between the layers remains anti-parallel, which does not necessitate any SET operation. System-level power-aware design methodologies can exploit this understanding to bypass any unnecessary SET operations.~\cite{pedra02} For NOR logic, such advantage is marginal. 

It needs to be emphasized that the inputs and output of the proposed straintronic logic gates are electrical in nature and hence there is no spin-to-charge conversion issue. Universal logic gates can be concatenated to perform any Boolean logic operation so any logic functionality can be implemented. However, unprecedented use of connectivity between the 2-input universal logic gates is not recommended, e.g., a majority logic gate or a 3-input logic gate may be required for different purposes in digital integrated circuits,~\cite{rabae03} which can be implemented following the same methodology of using a \emph{single} multiferroic element. 

In conclusion, we have devised a logic design concept utilizing \emph{single} multiferroic composites for the purposes of room-temperature computing that can be so energy-efficient that it can be powered from energy harvested from the environment. The basic building blocks are fast in operation, non-volatile (that can lead to instant turn-on computer), and they portend highly-dense logical functionality per unit area because of using \emph{single} multiferroic elements as universal logic gates. The proposed methodology is verified with an widely accepted model and it is within the reach of experimental implementation. Processors based on this paradigm can harbinger unprecedented applications such as medically implanted devices monitoring epileptic patient's brain to warn an impending seizure by drawing energy solely from the patient's body movements, or even energy radiated by wireless networks and television stations.

%\bibliographystyle{aipnum4-1}
%\bibliography{royk,royk2}
%merlin.mbs aipnum4-1.bst 2010-07-25 4.21a (PWD, AO, DPC) hacked
%Control: key (0)
%Control: author (8) initials jnrlst
%Control: editor formatted (1) identically to author
%Control: production of article title (-1) disabled
%Control: page (0) single
%Control: year (1) truncated
%Control: production of eprint (0) enabled
%

\end{document}

\preprint{AIP/123-QED}

\title{Supplementary Information\\Ultra-low-energy non-volatile straintronic computing using single multiferroic composites}% Force line breaks with \\

\author{Kuntal Roy}
\email{royk@vcu.edu.}
\noaffiliation
\affiliation{School of Electrical and Computer Engineering, Virginia Commonwealth University, Richmond, VA 23284, USA}
\thanks{Current affiliation: School of Electrical and Computer Engineering, Purdue University, West Lafayette, IN 47907, USA.}

%\date{\today}% It is always \today, today,
             %  but any date may be explicitly specified

\maketitle

\begin{figure}[h]
\centering
\includegraphics[width=70mm]{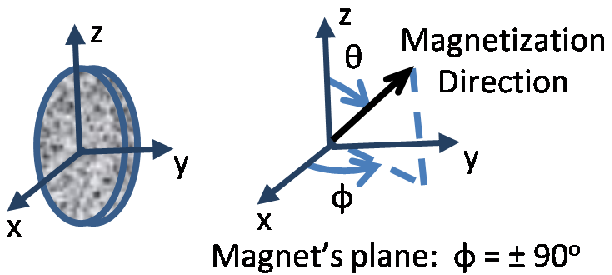}
\caption{\label{fig:magnetization_magnet_coordinates} {Magnetization in three-dimensional space.} Standard spherical coordinate system is used with $\theta$ being the polar angle and $\phi$ being the azimuthal angle. A uniaxial stress is generated along the easy axis ($z$-axis) through a piezoelectric layer via $d_{31}$ coupling.~\cite{roy11_6}}
\end{figure}

\begin{figure*}[h]
\centering
\includegraphics[width=170mm]{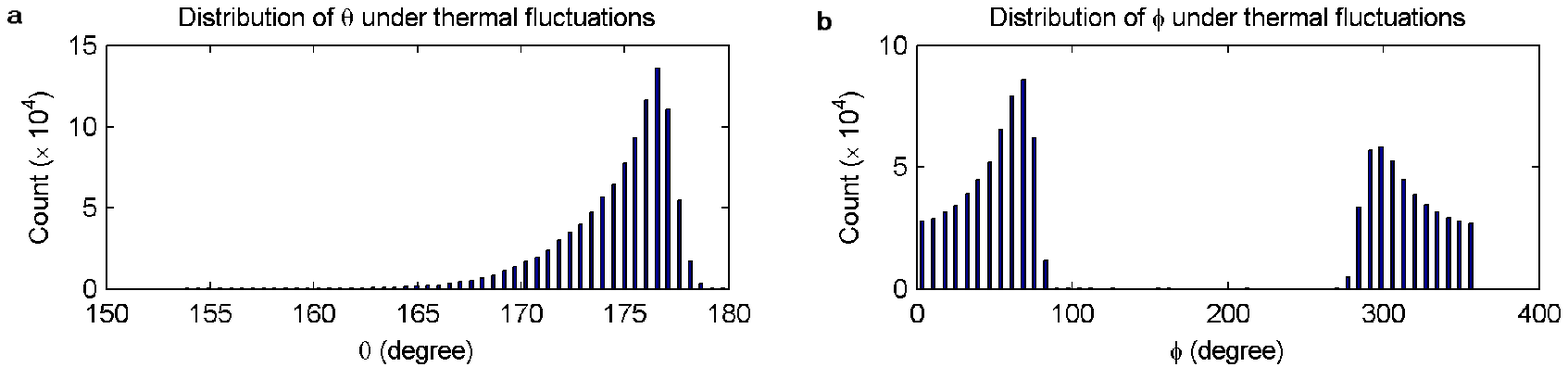}
\caption{\label{fig:NOR_initial_dist_theta_phi} Distribution of initial orientation of magnetization due to thermal fluctuations at room temperature (300 K) for the nanomagnet used to design the NOR gate. Distributions of both the polar angle $\theta_{initial}$ and the azimuthal angle $\phi_{initial}$ are plotted. A bias field of flux density 40 mT is applied along the out-of-plane hard axis (+$x$-direction), the sole purpose of which is to deflect the peak of magnetization's distribution exactly from the easy axis $\theta=180^\circ$.~\cite{roy11_6}
(a) Distribution of polar angle $\theta_{initial}$ at room temperature (300 K). The very initial position of the magnetization while starting the simulation was at $\theta=180^\circ$. The mean of the distribution is $174.7^\circ$, and the most likely value is 176.6$^{\circ}$.
(b) Distribution of the azimuthal angle $\phi_{initial}$ due to thermal fluctuations at room temperature (300 K). There are two distributions with  peaks centered at 68.4$^\circ$ and 298.8$^\circ$. The reason behind the asymmetry between these two distributions is explained in the Ref.~\onlinecite{roy11_6}.}
\end{figure*}

\begin{figure*}[h]
\centering
\includegraphics[width=170mm]{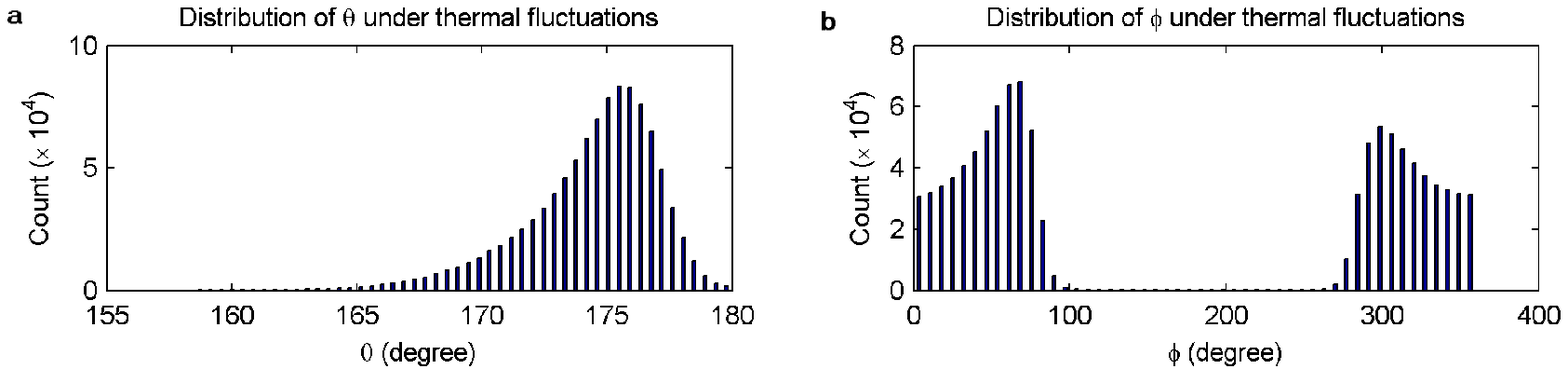}
\caption{\label{fig:NAND_initial_dist_theta_phi} Distribution of initial orientation of magnetization due to thermal fluctuations at room temperature (300 K) for the nanomagnet used to design the NAND gate. Distributions of both the polar angle $\theta_{initial}$ and the azimuthal angle $\phi_{initial}$ are plotted. A bias field of flux density 40 mT is applied along the out-of-plane hard axis (+$x$-direction), the sole purpose of which is to deflect the peak of magnetization's distribution exactly from the easy axis $\theta=180^\circ$.~\cite{roy11_6}
(a) Distribution of polar angle $\theta_{initial}$ at room temperature (300 K). The very initial position of the magnetization while starting the simulation was at $\theta=180^\circ$. The mean of the distribution is $174.4^\circ$, and the most likely value is 175.5$^{\circ}$.
(b) Distribution of the azimuthal angle $\phi_{initial}$ due to thermal fluctuations at room temperature (300 K). There are two distributions with  peaks centered at 68.4$^\circ$ and 298.8$^\circ$. The reason behind the asymmetry between these two distributions is explained in the Ref.~\onlinecite{roy11_6}.}
\end{figure*}

\textit{Distributions of initial orientations of magnetization due to thermal fluctuations for the nanomagnets used to design the universal logic gates.--} We adopt the standard spherical coordinate system (see Fig.~\ref{fig:magnetization_magnet_coordinates}) to solve the magnetization dynamics using stochastic Landau-Lifshitz-Gilbert (LLG) equation in the presence of room-temperature (300 K) thermal fluctuations.~\cite{roy11_6} We determine the initial distributions of polar angle $\theta$ and azimuthal angle $\phi$ at room-temperature by solving the stochastic LLG equation when no stress is active.\cite{roy11_6} Figures~\ref{fig:NOR_initial_dist_theta_phi} and~\ref{fig:NAND_initial_dist_theta_phi} plot the initial distributions $\theta_{initial}$ and $\phi_{initial}$ at room-temperature for the nanomagnets considered for designing the NOR gate and the NAND gate, respectively. 

\makeatletter 
\renewcommand\@biblabel[1]{$^{S#1}$}
\makeatother

%\vspace*{5mm}
%\noindent\textbf{References}\vspace*{-2mm}
%\bibliographystyle{aipnum4-1}
%\bibliography{royk,royk2}% Produces the bibliography via BibTeX.

%merlin.mbs aipnum4-1.bst 2010-07-25 4.21a (PWD, AO, DPC) hacked
%Control: key (0)
%Control: author (8) initials jnrlst
%Control: editor formatted (1) identically to author
%Control: production of article title (-1) disabled
%Control: page (0) single
%Control: year (1) truncated
%Control: production of eprint (0) enabled
%